# Quantum Spin Hall Effect and Topological Phase Transition in Two-Dimensional Square Transition Metal Dichalcogenides

Yandong Ma[†][*], Liangzhi Kou[‡], Ying Dai[§], and Thomas Heine[†][*]

[†] Engineering and Science, Jacobs University Bremen, Campus Ring 1, 28759 Bremen, Germany

[‡] Integrated Materials Design Centre (IMDC), School of Chemical Engineering, University of New South Wales, Sydney, NSW 2052, Australia

[§] School of Physics, Shandong University, 250100 Jinan, People's Republic of China

*Corresponding author: myd1987@gmail.com (Y.M.); t.heine@jacobs-university.de (T.H.)

Two-dimensional (2D) topological insulators (TIs) hold promise for applications in spintronics based on the fact that the propagation direction of edge electrons of a 2D TI is robustly linked to their spin origination. Here, with the use of first-principles calculations, we predict a family of robust 2D TIs in monolayer square transition metal dichalcogenides ($MoS_2$, $MoSe_2$, $MoTe_2$, $WS_2$, $WSe_2$, and $WTe_2$). Sizeable intrinsic nontrivial bulk band gaps ranging from 24 to 187 meV are obtained, guarantying the quantum spin Hall (QSH) effect observable at room temperature in these new 2D TIs. Significantly different from most known 2D TIs with comparable band gaps, these sizeable energy gaps originate from the strong spin-orbit interaction related to the pure d electrons of the Mo/W atoms around the Fermi level. A single pair of topologically protected helical edge states is established for the edge of these systems with the Dirac point locating in the middle of the bulk band gap, and their topologically nontrivial states are also confirmed with nontrivial topological invariant $Z_2 = 1$. More interestingly, by controlling the applied strain, a topological quantum phase transition between a QSH phase and a metallic phase or a trivial insulating phase can be realized in these 2D materials, and the detailed topological phase diagram is established.

TOC Figure

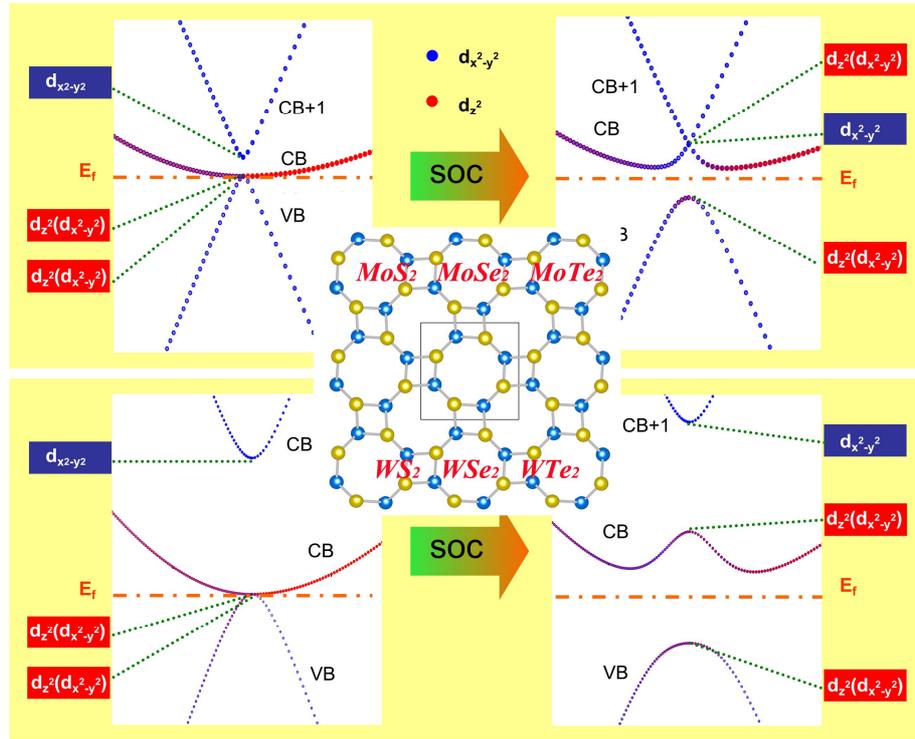

## I. Introduction

As the rise of two-dimensional (2D) materials, transition-metal dichalcogenides (TMDC, generally with phase of 1H and 1T), possessing diverse electronic and magnetic properties, have attracted renewed research interest, owing to their layered structures resembling graphite.[1-4] They can exhibit semiconducting, metallic and even superconducting behaviors depending on the combination of chalcogen and transition metal. $MX_2$ (M=Mo, W; X=S, Se, Te) are typical examples of the layered TMDC family. A monolayer of $MX_2$ is a chemically stable 2D material similar to graphene.[1,2] It is composed of three atomic layers, a hexagonal layers of M atoms sandwiched between two layers of X atoms. The sandwich layer is tightly bound internally and interacts with neighboring sandwich layers only through weak van der Waals (vdWs) interaction. Unlike semimetallic graphene, pristine monolayers of $MX_2$ are direct band gap semiconductors with band gap values ranging from 1.1 to 1.9 eV.[5-9] With relative fabrication easiness, chemical stability, relatively high mobility, and strong spin-orbit coupling (SOC), these materials are expected to have a significant impact on next-generation ultrathin electronic, optoelectronic, and valleytronic devices.[10-13]

Following the known theoretical verification of quantum spin Hall (QSH) effect in 2D topological insulators (TIs) graphene,[14] extensive effect has been devoted to the search for new 2D materials or new schemes to obtain 2D TIs with large band gaps.[15-19] Interestingly, it was shown recently that monolayer $MX_2$ can become 2D TIs for realizing QSH effect when the materials transform into 1T′ structure.[20] Generally the frontier orbitals of monolayer $MX_2$ are dominated by the d orbitals irrespective of the 1H or 1T structure, which show normal band order, thus resulting in the trivial topological phase in monolayer $MX_2$. While for the 1T′ structure, the states around the Fermi level are mainly contributed by the p and d orbitals. Such structural distortion in monolayer $MX_2$ with 1T′ structure leads to an intrinsic band inversion between X-p and M-d bands.[20] In these materials, the spin-filtered edge states exhibit dissipationless spin and charge transport which are immune to the nonmagnetic scattering, thus supporting promising applications in spintronics and quantum computations.[21]

In this work, we report a series of nontrivial 2D TIs with a sizeable band gap in square phase of monolayer $MX_2$ (defined as 1S-$MX_2$) in terms of first-principles calculations. The calculated $Z_2$ invariants and edge states provide direct evidence for their nontrivial topological characteristic. These new 2D TIs identified here all have relative sizeable energy gaps that exceed the thermal energy at room temperature, making these materials intriguing for applications at room-temperature. Different from the 1T′ structure-based TIs where the frontier orbitals are mainly from the p and d orbitals,[20] the frontier orbitals of 1S phase are dominated solely by the d electrons. When the strain is applied, the electronic state experiences a significant change at a certain strain range. For monolayer $MoTe_2$, the QSH phase can be retained and its topological gap can be turned. While for monolayer $MoS_2$, $MoSe_2$, $MoTe_2$, $WS_2$ and $WSe_2$, there is a topological quantum phase transition between a QSH phase and a metallic phase or a trivial insulating phase. The effectiveness of strain modified topological phase renders them more attractive for applications in semiconductor industry.

**II. Computational Methods**

Our calculations are based on density functional theory using the generalized gradient approximations (GGA)[22] of Perdew–Burke–Ernzerhof (PBE)[23] for electron-electron interactions, as implemented in Vienna *ab initio* simulation pack (VASP) code[24,25]. The projector augmented wave (PAW) method[26,27] is selected in the DFT calculations. The periodic boundary condition is used to simulate the monolayer $MX_2$. The vacuum space between two layers is set to 18 Å to avoid spurious interactions between periodic images. The Brillouin zone integration is performed with a 9×9×1 and a 17×17×1 k mesh for geometry optimization and self-consistent electronic structure calculations, respectively. An energy cutoff of 500 eV is used for the plane-wave expansion of the electronic wave function. Geometry structures are fully relaxed until the force on each atom is less than 0.01 eV/Å, and the convergence criteria for energy is $10^{-6}$ eV. Phonon dispersion relations are obtained with the finite displacement method using the CASTEP code[28,29], with the exchange correlation energy also being described by the functional of PBE.

## III. Results and Discussion

**Fig. 1a** shows the typical optimized lattice structure for $1S-MX_2$. Similar to the commonly studied monolayer $1H-MX_2$, $1S-MX_2$ can be also viewed as a three-layer stacking of M and X atoms, wherein M atoms are sandwiched between layers of X atoms and each M atom is coordinated to six X atoms.[30] Instead of the six-membered rings, they have been shown to possess interesting four- and eight-membered rings which were found at the grain boundary structure in $1H-MX_2$[31]. In this configuration, as shown in **Fig. 1a**, the square-octagon pairs are repeated along the *a* and *b* axis to form a 2D sheet. The optimized crystal structure of $1S-MX_2$ presents a square Bravais lattice with *p*4 symmetry and with four M and eight X atoms in one unit cell (marked by the black lines in **Fig. 1a**). The inversion symmetry holds for all these compounds. The equilibrium lattice constants are 6.336, 6.613, 7.055, 6.359, 6.643, and 7.105 Å for $1S-MoS_2$, $1S-MoSe_2$, $1S-MoTe_2$, $1S-WS_2$, $1S-WSe_2$, and $1S-WTe_2$, respectively. In **Fig. S1** we present the calculated the phonon dispersion of $1S-MX_2$ to confirm their stability. It is seen that the frequencies of all phonon branches in the whole Brillouin zone have positive values. This shows that these compounds are stable, corresponding to the energy minimum in the potential energy surface.

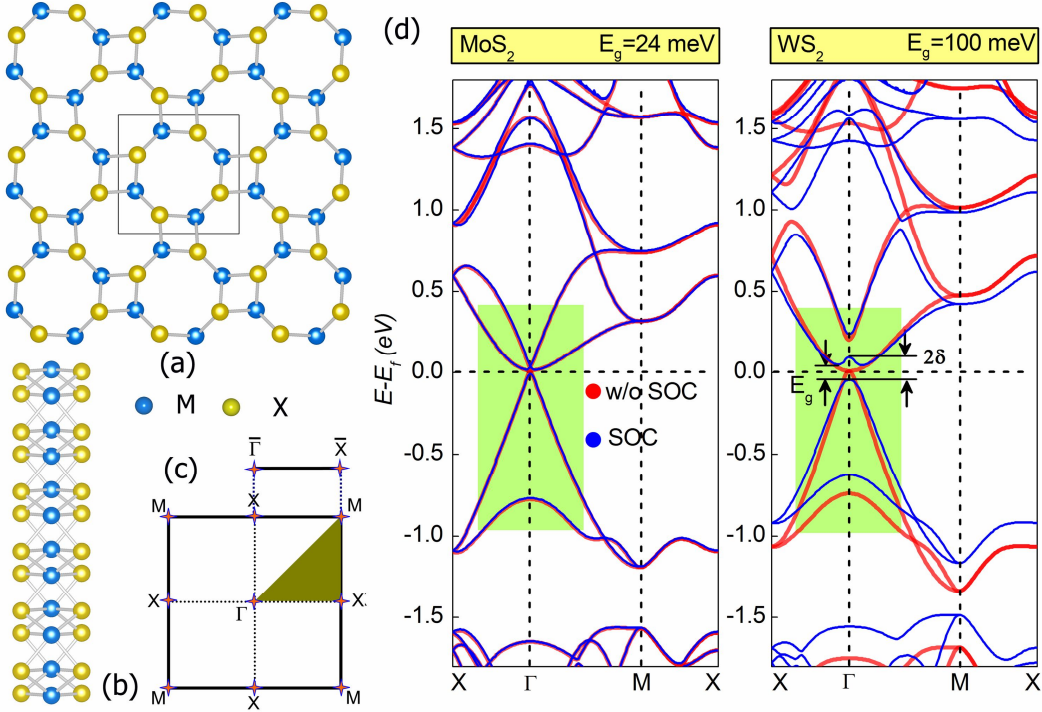

**Figure 1**. Crystal structures of monolayer 1S-MX$_2$ from (a) top and (b) side views. (c) 2D and projected one-dimensional (1D) Brillouin zones with high symmetry points. (d) Band structures of monolayer 1S-MoS$_2$ and 1S-WS$_2$ with and without SOC. The Fermi level is set to zero.

Typical band structures of 1S-MoS$_2$ and 1S-WS$_2$ are shown in **Fig. 1d**, and the corresponding detailed bands around the Fermi level are depicted in **Fig. 2a**. The band structures of other monolayers are shown in **Fig. S2**. In the case without SOC, the valence band maximum (VBM) and conduction band minimum (CBM) of all these monolayers meet in a single point at the Γ point, with the Fermi level locating exactly at the touching points. These monolayers can thus be regards as gapless semiconductors or semi-metals with zero density of states at the Fermi level. Therefore, the electronic characters of 1S-MX$_2$ are different from those of monolayers 1H-MX$_2$ which are semiconductors possessing a direct band gap with VBM and CBM situated at the K point, although they share the same composition. From the 1H configuration with p6m symmetry to the 1S-MX$_2$ with p4 symmetry, some structure symmetries are lost. From the view of the orbital-resolved band structures of 1S-MoS$_2$ and 1S-WS$_2$ plotted in **Fig. 2a**, the highest valence band and lowest conduction band (VB and CB) of 1S-MX$_2$ are mainly contributed by the $d_{x2-y2}$ (blue) and $d_{z2}$ (red) orbitals of the M atoms, respectively, with almost negligibly contributions from the p orbitals of X atoms. Notably, the hybridization between VB and CB of these monolayers leads to that the touching points (i.e., VBM and CBM) are dominated by mixed orbitals of $d_{x2-y2}$ and $d_{z2}$. Interestingly, for 1S-MoS$_2$, the CB+1 at the Γ point also approaches the degenerated point of VBM and CBM, locating 3 meV above the touching point. Such electronic characteristic would have a slight affect on the nontrivial fundamental band gap of 1S-MoS$_2$, as we will show below. To gain a deeper insight into the dispersion relation of 1S-MoS$_2$ around the meeting point, the corresponding bands around the Fermi level in three dimensions are presented in **Fig. 2b**. It can be seen that the three bands behave like a Dirac-like cones with lower and upper cones separated by a concave surface. Unlike the case of 1S-MoS$_2$, for the other five monolayers studied here, the CB+1 is well separated from the meeting points, see **Fig. 1d** and **Fig. S2**.

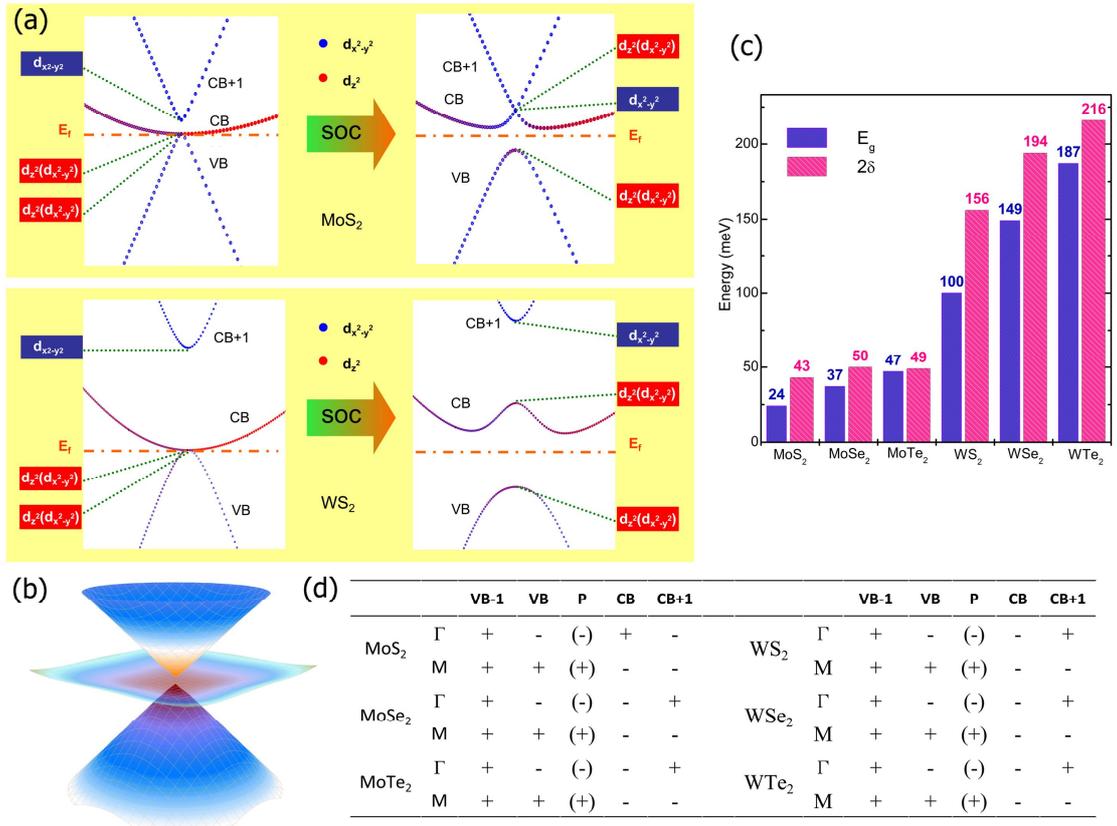

**Figure 2**. (a) The evolution of orbital-resolved band structures of 1S-MoS$_2$ and 1S-WS$_2$ with SOC. (b) Band structure of 1S-MoS$_2$ around the Fermi level in three dimensions. (c) The fundamental band gap ($E_g$) and the SOC-induced band gap at the Γ point (2δ) of 1S-MX$_2$. (d) The parities of

VB-1, VB, CB and CB+1 at the Γ and M points for 1S-MX$_2$; the products of the occupied bands at the Γ and M points are also listed in the brackets.

By turning on SOC, a metal-to-insulator transition occurs in 1S-MX$_2$, as the degenerated states at the meeting points separate from each other; see **Fig. 1d** and **Fig. S2**. This opening of band gaps strongly indicates that these monolayers may be 2D TIs. The SOC-induced splitting between the degenerated states at the Γ point are 43, 50, 49, 156, 194, and 216 meV, respectively, for 1S-MoS$_2$, 1S-MoSe$_2$, 1S-MoTe$_2$, 1S-WS$_2$, 1S-WSe$_2$, and 1S-WTe$_2$. We label this SOC-induced band gap as 2δ, to be distinguished from the fundamental band gap (E$_g$). Regarding the value of 2δ shown in **Fig. 2c**, we can see the SOC-induced band gap 2δ of 1S-WX$_2$ is significantly larger than that of the corresponding 1S-MoX$_2$. This is due to the stronger SOC strength in 5d electrons of the W atoms with respect to the 4d electrons of the Mo atoms. By projecting the bands onto different atomic orbitals, we find that for 1S-MoS$_2$, the VBM at the Γ point is contributed by the mixed orbitals of d$_{x2-y2}$ and d$_{z2}$, while its CBM at the Γ point is mainly from d$_{x2-y2}$ orbitals, as shown in **Fig. 2a**. In contrast to 1S-MoS$_2$, for the other five monolayers, both the VBM and CBM at the Γ point are dominated by the mixed orbitals of d$_{x2-y2}$ and d$_{z2}$. Such a discrepancy is sought in the band structures without considering SOC. As we mentioned above, unlike the other monolayers where the CBM+1 is well separated from the meeting point, without including SOC, the CBM+1 of 1S-MoS$_2$ locates only 3 meV above the touching point. Therefore, when turning on SOC, a band inversion between the CBM and CBM+1 occurs at the Γ point. While for the other monolayers, no such band inversion between the CBM and CBM+1 is observed. Note that the CBM+1 at the Γ point mainly originates from d$_{x2-y2}$ orbitals, this discrepancy can be easily understood. Since the CBM of 1S-MX$_2$ slightly shifts off the Γ point in the presence of SOC, the fundamental band gap of these monolayers would be smaller than their 2δ, ranging from 24 to 187 meV. Especially for 1S-WS$_2$, 1S-WSe$_2$, and 1S-WTe$_2$, the nontrivial fundamental band gaps are as large as 100, 149, and 187 meV, respectively, which far exceed the thermal energy at room temperature. Such sizeable band gaps of 1S-MX$_2$ is due to the strong SOC within the d electrons of the M atoms. It would make these monolayers being stabilized against considerable crystal effect and thermal fluctuation, which is essential for applications in high-temperature spintronics.

To verify the nontrivial topological nature of 1S-MX$_2$, we calculate the $Z_2$ topological invariants within the first-principles frameworks as direct evidence. In two dimensions, the band topology is given by one $Z_2$ topological invariant ($v$), with $v$=1 characterizing a topologically nontrivial phase and $v$=0 meaning a topologically trivial phase. Due to the fact that inversion symmetry holds for all these monolayers studied here, the calculation of topological invariant is expected to be simplified. Following the method developed by Fu and Kane[32], it can be directly obtained from the knowledge about the parity of each pair of Kramer's degenerate occupied energy band at the time-reversal-invariant momenta (TRIM). The Brillouin zone of 1S-MX$_2$ is shown in **Fig. 1c**, which is square with four X points on the side centers and four M points on the coiners. And there are four TRIM points for 1S-MX$_2$: one Γ point, one M point and two X points. The $Z_2$ topological invariants ν for 1S-MX$_2$ are thus given by,

$$\delta(K_i) = \prod_{m=1}^{N} \xi_{2m}^i, \quad (-1)^v = \prod_{i=1}^{4} \delta(K_i) = \delta(\Gamma)\delta(M)\delta(X)^2.$$

Herein, the $\delta(K_i)$ presents the product of parity eigenvalues at the TRIM points, $\xi=\pm1$ are the parity eigenvalues and $N$ is the number of the degenerate occupied energy bands. The corresponding results are listed in **Fig. 2d**. We find that 1S-MX$_2$ displays a nontrivial band topology with $Z_2$ topological invariant $v$=1. Consequently, all these monolayers are indeed nontrivial 2D TIs. Considering their sizable nontrivial band gaps, the QSH effect can be readily realized in all these systems. To gain further insight into the underlying mechanism for the obtained TI behaviors, we study the electronic states around the Fermi level at the Γ point. In the absence of SOC, the hybridization between VB and CB leads to the formation of a single point at the Γ point, with the two degenerated states both displaying "-" parities (spin degeneracy is neglected here). When turning on SOC, the degeneracy of this pair of "-" parities is lifted, with one shifting downward and another upward. During this process, no parity exchange between the valence and conduction bands is observed. Therefore, the inclusion of SOC could not induce any band inversion between the valence and conduction bands at the TRIM points in 1S-MX$_2$; instead, the only effect of SOC in these systems is rather to open an energy gap at the meeting point afterwards. This is different from the TIs where SOC is responsible for the band inversion. Such characteristic of "intrinsic nontrivial band order" was also found in several other systems, such as the chemically modified Bi and Sn honeycomb lattices.[15,16,19]

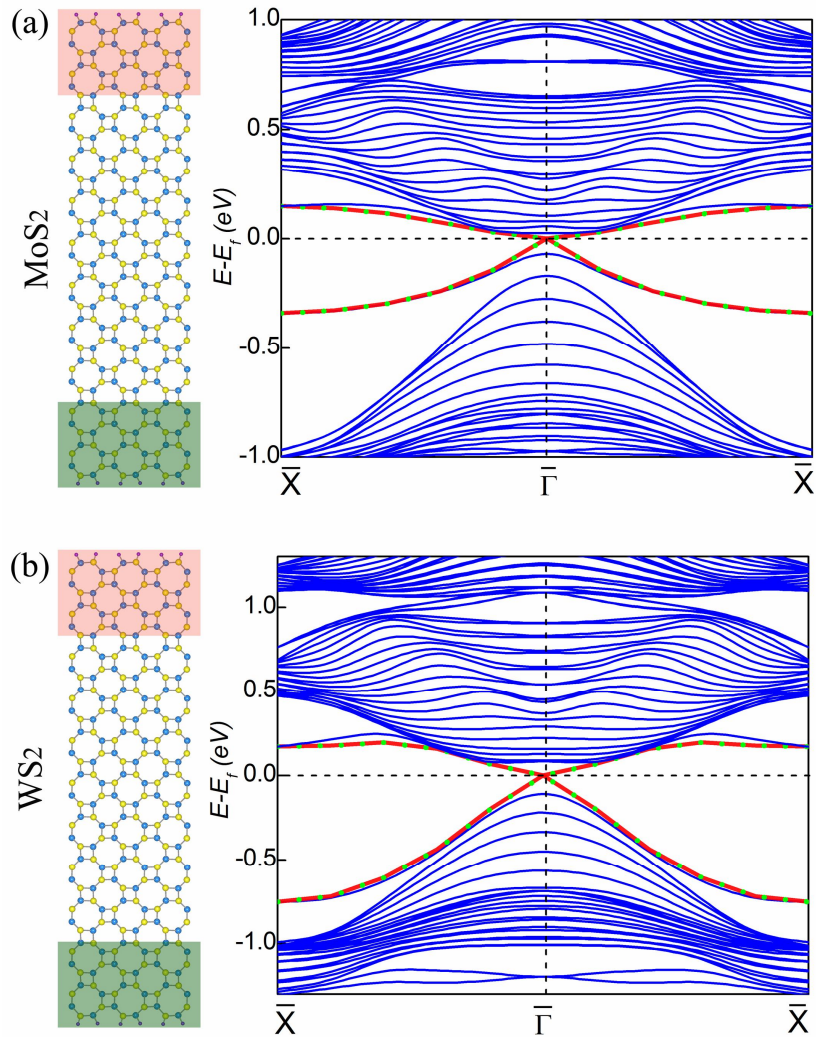

**Figure 3**. Crystal and band structures of (a) 1S-MoS$_2$ and (b) 1S-WS$_2$ ribbons. The bands contributed from the top and bottom edges are marked with red lines and green circles, respectively. The Fermi level is set to zero.

The 2D nontrivial insulating state in 1S-MX$_2$ should support an odd number of topologically protected conducting edge states connecting the conduction and valence bands. The edges of 1S-MX$_2$ are introduced by constructing a nanoribbon with all the dangling bonds passivated by the hydrogen atoms. To avoid the interaction between two edges, we adopt a 11-unit-cell-thick nanoribbon for 1S-MX$_2$ with width >6.7 nm. The calculated electronic structures of 1S-MoS$_2$ and 1S-WS$_2$ ribbons are presented in **Fig. 3**, while the corresponding results for the other systems are shown in **Fig. S3**. We can see that two pair of gapless edge states located at the opposite edges is present in the bulk band gap and they are energetically degenerated due to the symmetric edges of the nanoribbon. The red lines show the contribution from the top edge, while the green circles show the contribution from the bottom edge. It is important to notice that each edge state almost degenerates with the bulk bands near the $\bar{X}$ points (although such bulk bands looks like the edge states). The helical edge states (red or green) connect the bulk valence and conduction bands (blue) and cross at the $\bar{\Gamma}$ point, exhibiting the topologically nontrivial property. Such topologically protected edge states consistently confirm these monolayers are 2D TIs. It is also worth emphasizing that, thanks to the nontrivial topology of 1S-MX$_2$, the conducting edge states of these monolayers always exists irrespective of the type of the edge, but their details are affected by the atomic structures of the edges.

In the above discussion, we have firmly demonstrated the nontrivial topological properties with sizable bulk band gaps in 1S-MX$_2$. Yet, for the purpose of technological applications, it is obviously of particular interest to investigate the strain effect on the topological properties. This topic is of fundamental relevance science that strain may arise when a crystal is compressed or stretched out of equilibrium, which thus can significantly affect the device performance. On the other hand, sometimes strain is intentionally applied because advanced applications often require materials with topological properties which can be deliberately modulated in a well-controlled manner. Therefore, in the following, we will apply external strain on 1S-MX$_2$ to examine the modulation of the topological nature, and there are two main questions to be answered. One important question concerns whether the interesting topological phase transition can be induced in 1S-MX$_2$ by adjusting interatomic interaction. Another question of interest centers is whether—and to what extent—the nontrivial bulk band gap can be turned. Due to the structural isotropy, we impose strain on the in-plane xy direction of 1S-MX$_2$ equally by turning the planar lattice parameter, and for each fixed lattice parameter, the atomic positions are fully optimized. The strain magnitude is described by the quantity $\varepsilon = (a - a_0)/a_0$, where $a_0$ and $a$ is the lattice parameters of the unstrained and strained 1S-MX$_2$, respectively.

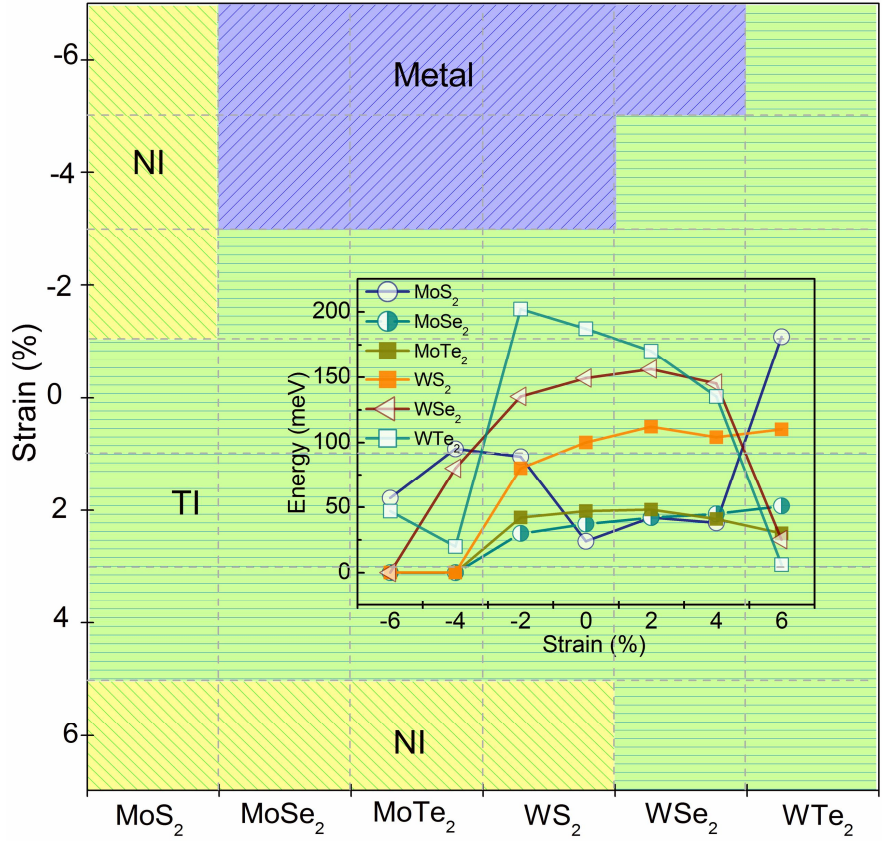

**Figure 4**. Topological phase diagram of 1S-MX$_2$ as a function of strain. The insert addresses the strain-dependent fundamental band gap (E$_g$) of 1S-MX$_2$ as a function of strain.

In **Fig. 4**, we show the topological phase diagram of 1S-MX$_2$ as a function of strain. And the detailed band structures of 1S-MX$_2$ under different strain are shown in **Fig. S4-S9**. It is seen that the topological phase is indeed sensitive to the applied strain, and topological phase transition can be induced in most of the monolayers. We find that 1S-MoS$_2$ is located very close to the boundary between a phase exhibiting the QSH effect and a trivial insulating phase. Within the strain range from 0% to 4%, 1S-MoS$_2$ is in the QSH phase, which is identified by the well-defined nontrivial Z$_2$ topological invariant. When the lattice is expanded by 6% or compressed by more than -2%, 1S-MoS$_2$ would transform in a trivial insulating phase, leading to a topological phase transition into 1S-MoS$_2$. While for 1S-MoSe$_2$, 1S-MoTe$_2$ and 1S-WS$_2$, their topological properties seem to response relatively slow to the imposed strain with respected to 1S-MoS$_2$. The largest compress strain can reach up to -2% without destroying the nontrivial topological behaviors in these three monolayers. When the compress strain reaches to -4%, a topological phase transition is also observed in 1S-MoSe$_2$, 1S-MoTe$_2$ and 1S-WS$_2$, however, they are transformed into a metallic phase instead of the trivial insulating phase. Therefore, with increasing strain from -6% to 6%, 1S-MoSe$_2$, 1S-MoTe$_2$ and 1S-WS$_2$ would undergo metallic phase → nontrivial topological phase → trivial insulating phase in sequence. Such diversified behaviors could provide us with an excellent opportunity to study the topological phase transition. While for 1S-WSe$_2$, as shown in **Fig. 4**, the nontrivial topological property is stable in a large strain range until the strain researches -6%. And at the strain of -6%, it is derived into the metallic phase, and no trivial insulating phase can be obtained in 1S-WSe$_2$ within the calculated strain range. This significant strain dependence demonstrates an important conclusion, that is, the nontrivial topological properties in these systems can be considerably altered by modifying the lattice. Bearing in mind these results, it will be very interesting to observe such topological phase transitions in these systems in experiments, simply by pressing or stretching the sample along the in-plane direction. The results have important technological implication. On the other hand, unlike these five systems, the nontrivial topological phase of 1S-WTe$_2$ survives in the strain range from -6% to 6% and no topological phase transition occurs, indicating its robust stability against the strains. This would make the QSH effect in 1S-WTe$_2$ highly adaptable in various application environments. Imposing external strain can not only induce interesting topological phase transition in these systems but also modify the nontrivial bulk band gaps in 1S-MX$_2$. The corresponding results are plotted in the insert in **Fig. 4**. It is worth highlighting that the nontrivial bulk band gaps of 1S-MX$_2$ can be tuned significantly by applying strain and that most of the nontrivial energy gaps remain larger than the thermal energy at room temperature. The comprehensive phase diagrams of 1S-MX$_2$ presented in **Fig. 4** would provide a tangible basis for guiding search of viable substrates for growing and realizing topological phase transitions in 2D-TI systems.

## IV. Conclusion

In conclusion, first-principles calculations have been used to study the electronic and topological properties of 1S-MX$_2$. All the monolayers studied here are predicted to be promising 2D TIs with nontrivial Z$_2$ topological invariants and a single pair of topologically protected helical edge states at the nanoribbon edge. These 2D TIs display sizable bulk band gaps ranging from 24 to 187 meV, which far exceed the thermal energy and could support observable QSH effect at room temperature. Moreover, the tunable electronic and topological properties of 1S-MX$_2$ with strain provide a feasible approach for band engineering and a basis for practical application. Most remarkably, a topological quantum phase transition between a phase exhibiting the QSH effect and a metallic phase or a trivial insulating phase can be realized in 1S-MX$_2$ by imposing external strain. These monolayers can provide a novel platform for diverse potential applications.

**Supporting Information**

Phonon band dispersion relations calculated for 1S-MX$_2$; electronic band structures for 1S-MoSe$_2$, 1S-MoTe$_2$, 1S-WSe$_2$, and 1S-WTe$_2$; band structures of 1S-MoSe$_2$, 1S-MoTe$_2$, 1S-WSe$_2$, and 1S-WTe$_2$ ribbons; band structures of 1S-MX$_2$ under different strains. This material is available free of charge via the Internet.

**Acknowledgement**

Financial support by the European Research Council (ERC, StG 256962) and the National Science foundation of China under Grant 11174180 are gratefully acknowledged.